\documentstyle[11pt,newpasp,twoside,epsf]{article}

\begin{document}

\title{Polarization of Gamma--Ray Burst Optical and Near-Infrared Afterglows}

\author{Stefano Covino, Gabriele Ghisellini}
\affil{INAF--Brera Astronomical Observatory, Merate, Italy}
\author{Davide Lazzati}
\affil{Institute of Astronomy, Cambridge, UK}
\author{Daniele Malesani}
\affil{INAF--Brera Astronomical Observatory,  Merate, Italy; \\
International School for Advanced Studies (SISSA--ISAS), Trieste, Italy}

\begin{abstract}
Gamma--Ray Burst afterglow polarization measurements, in spite of their
intrinsic difficulties, have been carried out for a number of events that
begins to be adequate to draw some general statistical conclusions. 
Although the presence of some
degree of intrinsic polarization seems to be well established, there are still
open problems regarding the polarization time evolution and the possible
contribution of polarization induced by dust in the host galaxies. 
\end{abstract}

\section{Why is polarimetry important for GRB afterglows?}

\begin{figure}
\plottwo{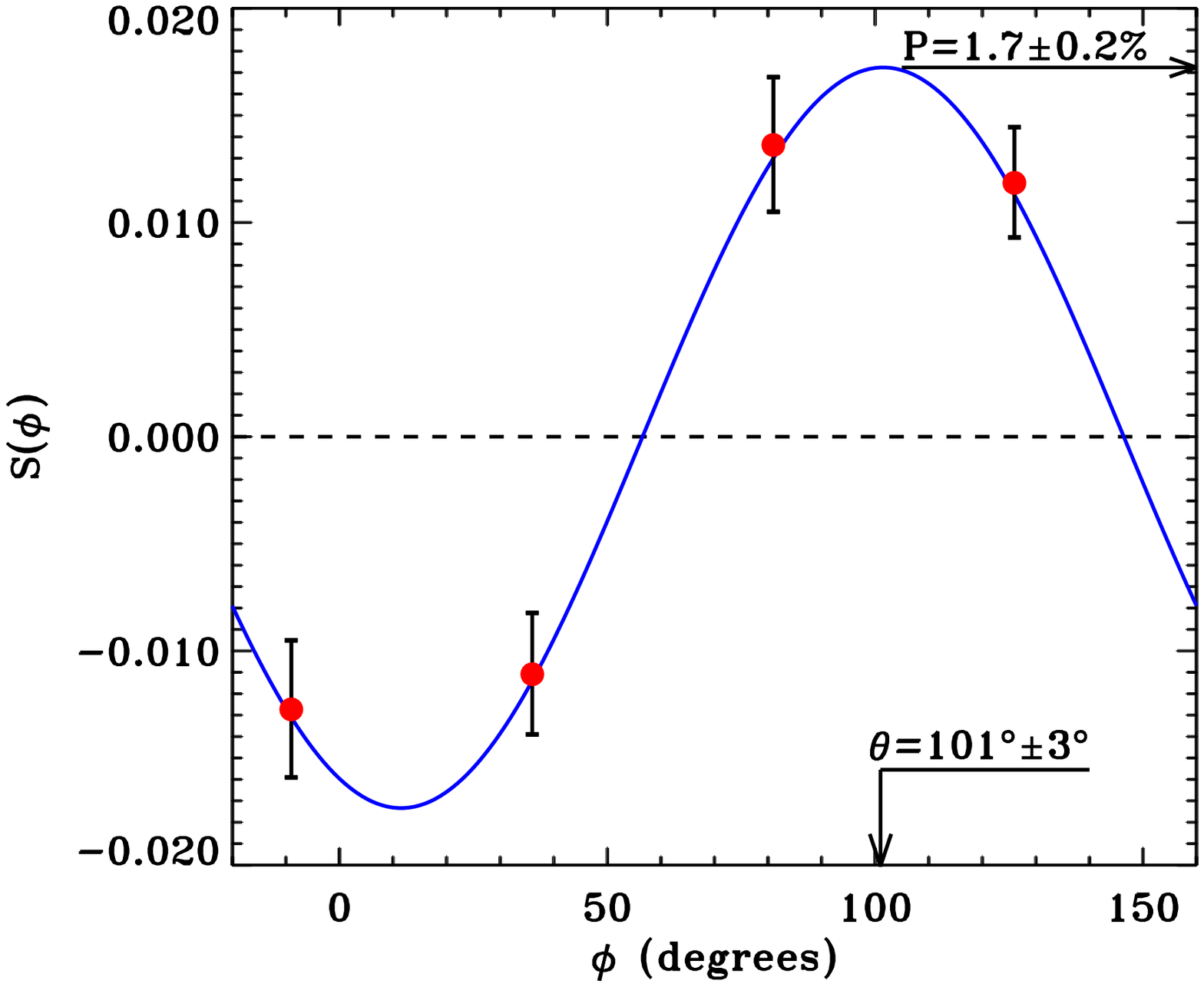}{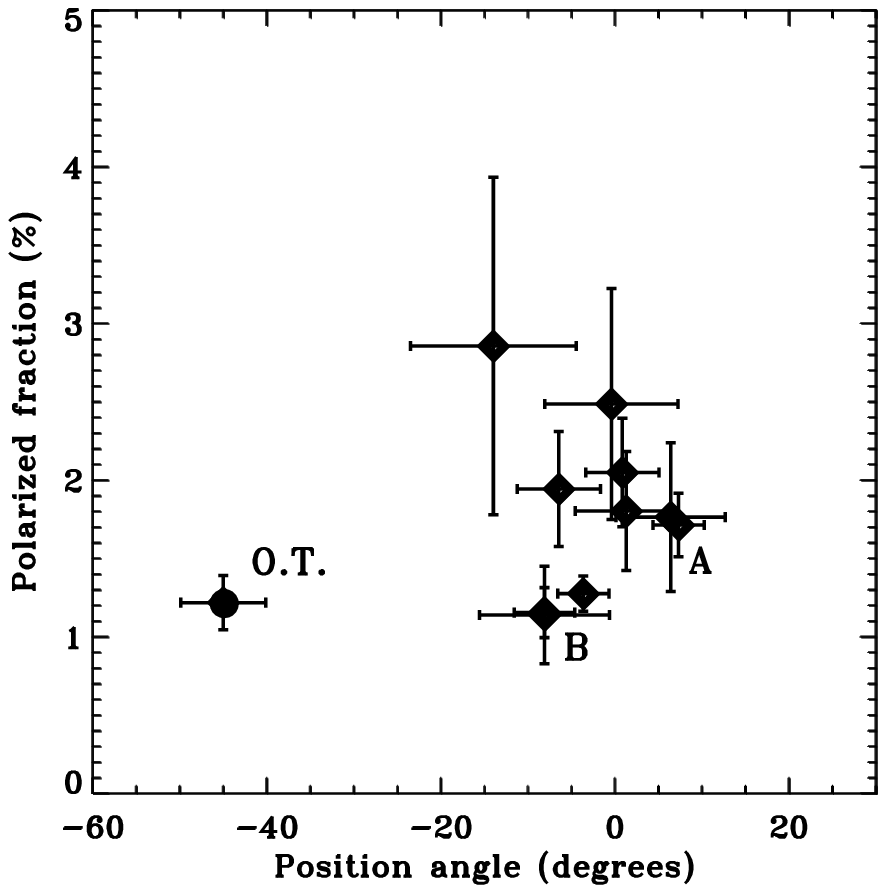}
\caption{{\it Left panel:} The net polarization of GRB\,990510.
{\it Right panel:}
Polarization level vs instrumental polarization position angle for the
stars in the field and the optical transient (not normalized). The optical
transient clearly stands out. From Covino et al. (1999).}
\label{fig:first}
\end{figure}

Polarization from astrophysical sources is a typical signature for a number of
physical phenomena (di Serego et al. 1997). In the context of gamma--ray burst
(GRB) physics, the simple detection of some degree of polarization from an
optical afterglow (Covino et al. 1999, Wijers et al. 1999; 
Fig.\,\ref{fig:first}) has always been considered a clear signature for
synchrotron emission. Time and wavelength variation of the polarization degree
and position angle can be powerful probes for the physics and the dynamics of
the expanding fireball and of the GRB environment.

\begin{table}[!t]
\caption{The results of the 27 polarization measurements performed so far 
(January 2003). $1 \sigma$ errors and  95\% confidence level upper limits are
reported. SP stands for spectropolarimetry. For GRB\,020813 and GRB\,021004 the
reported polarization degrees and position angles are not yet corrected for
Milky Way interstellar matter induced polarization.}
\begin{center}
\begin{tabular}{|l|ccccr|}
\hline
{\bf Burst} & {\bf $P$ (\%)} & {\bf $\vartheta$ ($^\circ$)} & {\bf $\Delta t$
(hours)} & {\bf Band} & {\bf Ref.} \\
\hline
GRB\,990123 & $<2.3$             &  & 18 & $R$ & 1 \\
GRB\,990510 & $1.7 \pm 0.2$ & $101 \pm 3$ & 18 & $R$ & 2 \\
                         & $1.6 \pm 0.2$ & $96 \pm 4$  & 21 & $R$ & 3 \\
			 & $< 3.9$     & &43      & $R$ &  3 \\
GRB\,990712 & $2.9 \pm 0.4$ & $122 \pm 4$ & 11 & $R$ & 4 \\
	 		 & $1.2 \pm 0.4$ & $116 \pm 10$ & 17 & $R$ & 4\\
			 & $2.2 \pm 0.7$ & $139 \pm 10$ & 35 & $R$ & 4\\
GRB\,991216 & $<2.7$  &  & 35 & $V$ & 5 \\
		     	 & $<5$  &   & 60 & $V$ & 5\\
GRB\,010222 & $1.4 \pm 0.6$ & & 22 & $V$ & 6  \\
GRB\,011211 & $< 2.0$ & & 37 & $R$ & 7 \\
GRB\,020405 & $1.5 \pm 0.4$ & $172 \pm 8$ & 29 & $R$ & 8 \\
			 & $9.8 \pm 1.3$ & $180 \pm 4$ & 31 & $V$ & 9 \\
			 & $2.0 \pm 0.3$ & $154 \pm 5$ & 52 & $V$ & 10 \\
			 & $1.5 \pm 0.4$ & $168 \pm 9$ & 78 & $V$ & 10\\
GRB\,020813 & $2.3 - 3.1$ & $153 - 162$ & 6 & SP & 11 \\
			 & $1.2 \pm 0.2$ & $158 \pm 5$ & 24 & $V$ & 12 \\
			 & $1.6 \pm 0.3$ & $163 \pm 6$ & 29 & $V$ & 13 \\
			 & $2.0 \pm 0.4$ & $179 \pm 6$ & 50 & $V$ & 13\\
			 & $4.3 \pm 1.7$ & $177 \pm 11$ & 96 & $V$ & 13\\
GRB\,021004 & $< 5$               & & 11 & $J$ & 5 \\
		  	 & $1.3 \pm 0.1$ & $114 \pm 2$ & 15 & $V$ & 14 \\
			 & $1.3 \pm 0.3$ & $125 \pm 1$ & 16 & $V$ & 15 \\
			 & $1.4 - 2.3$ & $111 - 126$ & 19 & SP & 16 \\
			 & $0.7 \pm 0.2$ & $89 \pm 10$ & 91 & $V$ & 17\\
\hline
\end{tabular}
\end{center}
1: Hjorth et al. 1999; 2: Covino et al. 1999; 3: Wijers et al. 1999;
4: Rol et al. 2000; 5: this paper; 6: Bj\"ornsson et al. 2002;
7: Covino et al. 2002d; 8: Masetti et al. 2002; 9: Bersier et al. 2003;
10: Covino et al. 2003a; 11: Barth et al. 2003; 12: Covino et al. 2002a;
13: Rol et al. 2003; 14: Covino et al. 2002b; 15: Rol et al. 2002;
16: Wang et al. 2003; 17: Covino et al. 2002c.
\label{tab:tabellona}
\end{table}

\subsection{How polarization can be produced in GRB afterglows}

\begin{figure}
\plotone{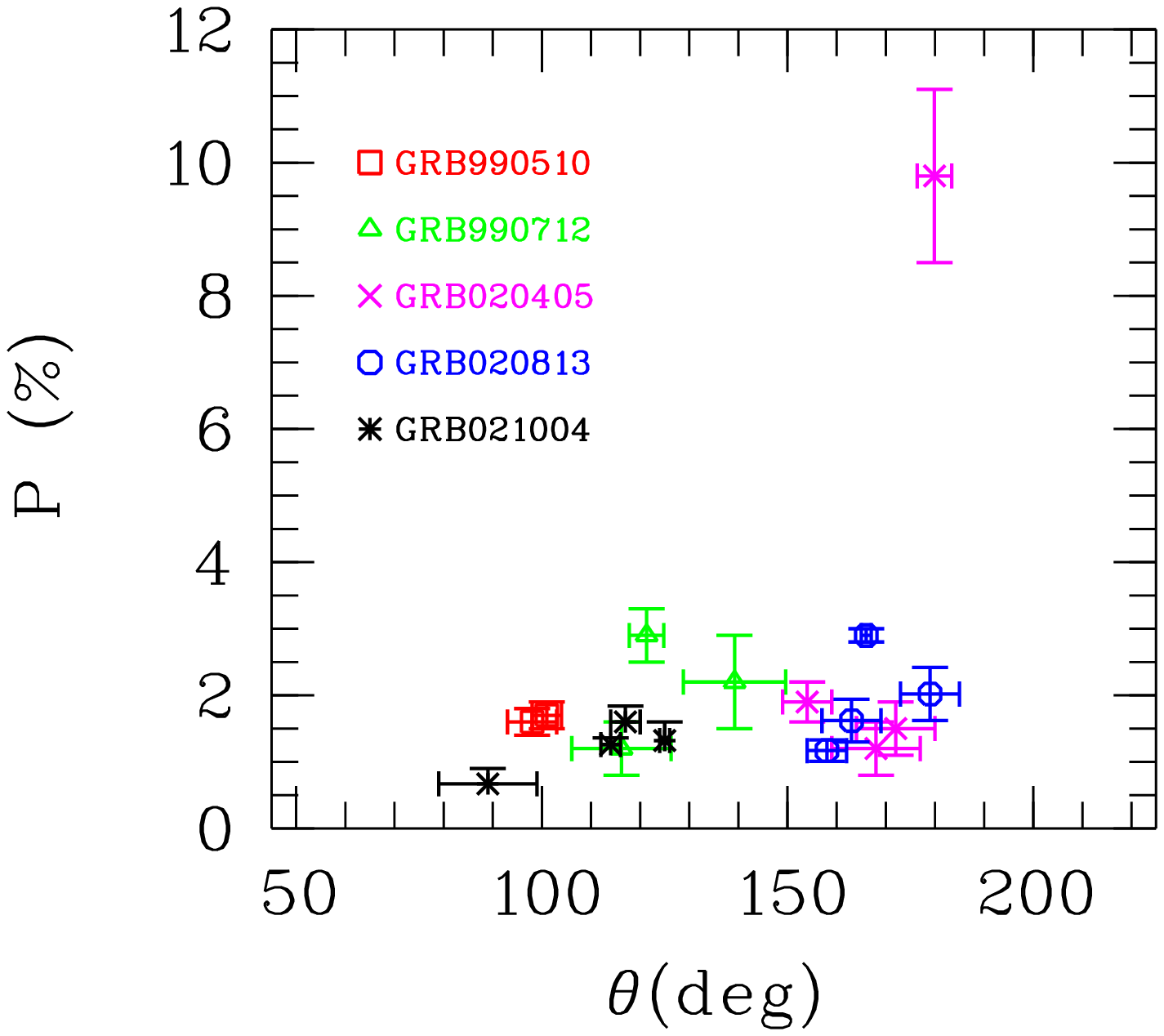}\\
\plotone{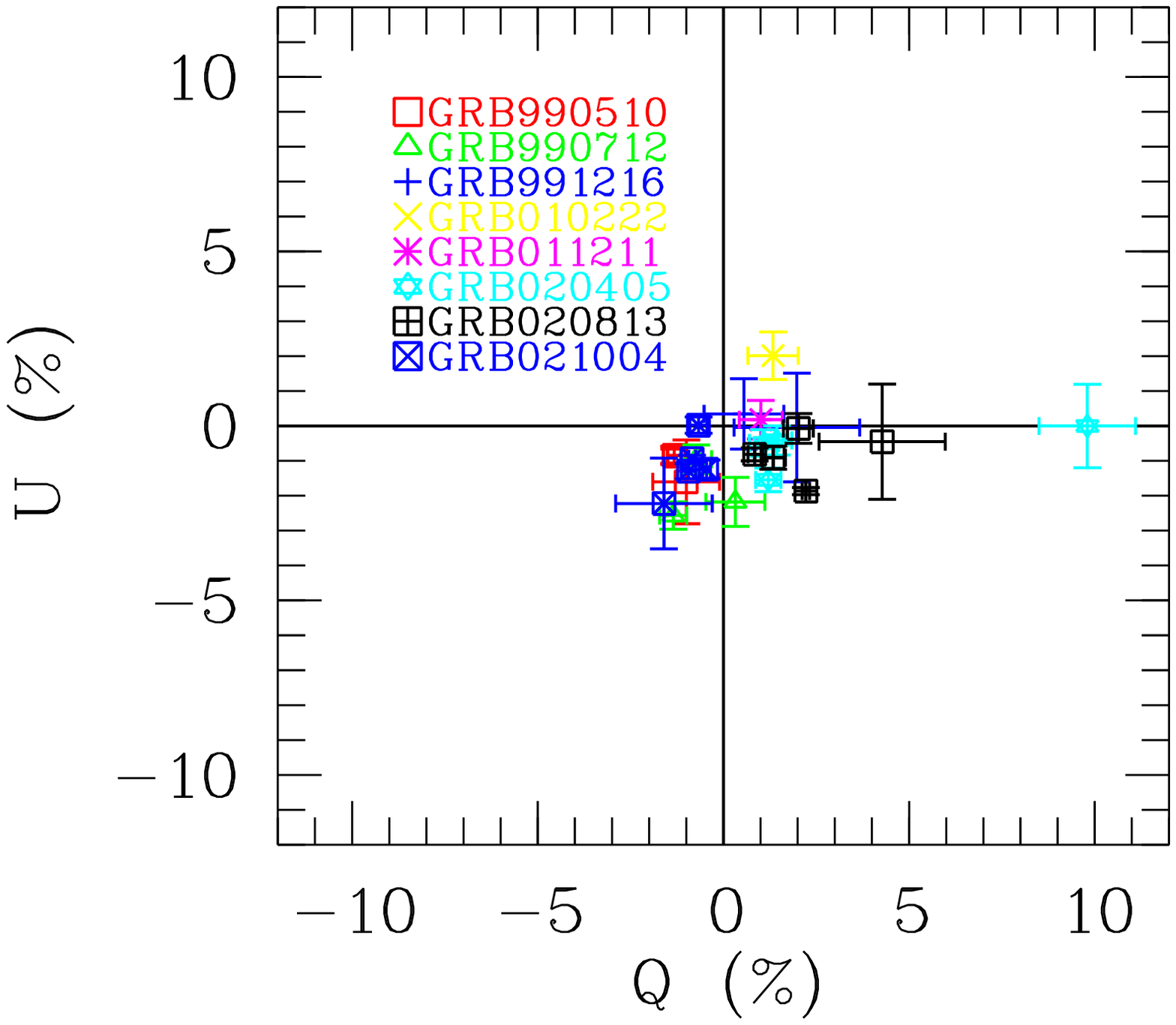} \\
\caption{{\it Top panel:} Polarization degree and position angle for all the
positive detections, i.e. upper limits are excluded. {\it Bottom panel:} $Q$
and $U$ Stokes' parameters for all the available data, i.e. including upper
limits.}
\label{fig:totpol}
\end{figure}

As a general rule, to produce observable polarization, some degree of 
anisotropy is necessarily required. Two general families of models have been 
developed to explain the level of polarization and its time evolution in GRB
optical afterglows. One possibility is that the emission originates in causally
disconnected regions of highly ordered magnetic field, each producing 
polarization almost at the maximum degree (Gruzinov \& Waxman 1999; Gruzinov
1999; Medvedev \& Loeb 1999; Loeb \& Perna 1998). The observed polarization is
then lowered by averaging over the unresolved source. Gruzinov \& Waxman (1999)
predicted a $\sim 10\%$ polarization. If the regions have a statistical 
distribution of energies, the position angle can be different at various
wavelengths. This value is somewhat greater than what has been observed in
GRB\,990510, GRB\,990712, GRB\,010222, GRB\,020405, GRB\,020813 and
GRB\,021004. It is also greater than the upper limits derived for GRB\,990123,
GRB\,990510, GRB\,991216 and GRB\,011211 (references in the caption of
Tab.\,\ref{tab:tabellona}). All positive detections so far derived are below
$\sim 3$\% (apart from one possible case, see Fig.\,\ref{fig:totpol}) while the
upper limits are lower than $\sim 5$\%.

In an alternative scenario proposed by Ghisellini \& Lazzati (1999), Sari
(1999), Granot et al. (2002), the magnetic field is ordered in the plane of the
shock. In a spherical fireball, such a field configuration would give null
polarization, but if a collimated fireball is observed off--axis (as it is most
probable), a small degree of polarization is predicted, with a well defined
time evolution.  The light curve of the polarized emission has two peaks, with
the position angle of the polarization vector shifting by 90$^\circ$ between
them. Another important feature of the model is that the ratio of the polarized
flux of the two peaks is related to the dynamics of the interaction of the jet
with the interstellar medium. The polarization position angle should be in this
case independent of the wavelength. Therefore, with a good time sampling of the
polarization intensity and position angle the geometry and the dynamics of the
jet can be efficiently pinned down.

Recently, Rossi et al. (2003; see also these proceedings) have predicted
the polarization arising in structured jets when observed off--axis (Rossi,
Lazzati \& Rees 2002; Zhang \& M\'esz\'aros 2002). Here the additional
asymmetry is provided by the fact that the energy carried by the jet per unit
solid angle is a function of the angle from the jet axis. In this model a
single peak of polarized emission is predicted, with a constant polarization
angle.

% Even if the main radiation process of the afterglows in various bands is
% synchrotron, there is still some debate about the possible occurrence of
% other
% processes (i.e. synchrotron self Compton or Comptonization; Meszaros \& Rees
% (1997); Ghisellini \& Celotti 1999).

A small contribution to the observed polarization by dust in the host galaxy
has not been completely ruled out. This could be the case if GRB progenitors
are located in dusty star--forming regions.  The key observational sequence to
investigate on these possibilities involves multiband polarization measurements
since the polarization induced by dust shows a well-defined wavelenght
dependance.  Determining the time and the wavelength behavior of the
polarization will therefore provide a wealth of new important information.

However, since polarization variability has been clearly singled out (e.g. in
GRB\,020813 and, possibly, in GRB\,020405), a significant fraction of the
polarization degree has to be intrinsic.

\section{Some historical steps for GRB polarization measurements}

Starting from 1999, at present (January 2003) 27 different polarization
measurements have been performed, by studying 10 different optical and near
infrared (NIR) afterglows. This resulted in 19 positive detections and 8 upper
limits.

The first attempt to measure polarization for a GRB afterglow was carried out
for GRB\,990123 by Hjorth et al. (1999). Although in this case only an upper
limit was derived (see Table\,\ref{tab:tabellona}), it was stringent enough to
pose severe constraints to the various emission models. The problem immediately
became not to explain how GRB optical afterglows (OA) can produce polarization,
but why the polarization degree was so small (or possibly zero). The first
successful polarization measurement came for the OA of GRB\,990510: Covino et
al. (1999) and Wijers et al. (1999) measured a polarization level at $\sim
1.7\%$, therefore supporting the synchrotron emission scenario for the
afterglows.

Few months later, Rol et al. (2000) were then able to carry out the first
successful set of multiple polarization measurements reporting a polarization
level around $2\%$ for more than 20 hours, with a constant position angle and a
marginal variation in intensity.

During the next couple of years several upper limits or marginal detections
were derived: for GRB\,991216 (this paper), GRB\,010222 (Bj\"orsson et al.
2002) and GRB\,011211 (Covino et al. 2002d). These findings again confirmed
that the polarization degree up to a couple of days after the burst is always
below few per cent. Attempts to measure polarization also in the NIR were
carried out by Klose et al. (2001), but they provided weakly constraining upper
limits of $\sim 30$\%.

In 2002, three sets of observations (for GRB 020405, GRB 020813 and GRB
011004), produced a wealth of new information which are partly still under
analysis. GRB 020405 was observed four times (Masetti et al. 2002, Bersier et
al. 2003, Covino et al. 2002a) and three of these observations again give a
roughly constant polarization degree ($1.5-2$\%) and position angle. On the
contrary, Bersier et al. (2003), two hours later than the measurement performed
by Masetti et al., estimated $\sim 10$\%. At present no model can explain such
a sharp variation, if real.

Eventually, GRB\,020813 and GRB\,021004 were observed five times each,
and spectropolarimetric observations were also derived (Barth et al. 2002; Wang
et al. 2003; this paper). Data analysis is still in progress but prelimimary
results show that the polarization degree and position angle seem to be
smoothly varying or roughly constant in the optical band. In both cases we hope
that when the final results for the whole data sets will be fully available,
the time coverage may be adequate enough to discrimate among different models
and to derive some important physical parameters.

\section{General considerations}

The polarimetric and spectropolarimetric observations performed so far may
allow to draw some statistical inference regarding polarization level and
position angle for the observed GRB OAs from few hours up to 2--3 days
after the $\gamma$--ray event. Fig. \ref{fig:totpol} shows the polarization
degree and position angle for all positive detections and the $Q$ and $U$
Stokes' parameters for all the performed observations (including upper limits).
Apart from one controversial and therefore potentially interesting case
(GRB\,020405, Bersier et al. 2003), all polarization measurements and upper
limits show a polarization degree roughly below $\sim 3$\%. Any possible model
has at least to be able to predict such a high probability to observe low or
null polarization in the emission of GRB OAs observed starting from a few hours
after the burst event.


\begin{references}
\reference Barth A.J., Sari R., Cohen M.H. et al. 2003, ApJ 584, L47
\reference Bersier D., McLeod B., Garnavich M. et al. 2003, ApJ 583, L63
\reference Bj\"ornsson G., Hjorth J., Pedersen K., Fynbo J.U. 2002, ApJ 579, 59
\reference Covino S., Ghisellini G., Malesani D. et al. 2002b, GCN\,1595
\reference Covino S., Ghisellini G., Malesani D. et al. 2002c, GCN\,1622
\reference Covino S., Lazzati D., Ghisellini G. et al. 1999, A\&A 348, L1
\reference Covino S., Lazzati D., Malesani D. et al. 2002d, A\&A 392, 865
\reference Covino S., Malesani D., Ghisellini G. et al. 2002a, GCN\,1498
\reference Covino S., Malesani D., Ghisellini G. et al. 2003, A\&A, in press
     (astro--ph/0211245)
\reference Hjorth J., Bj\"ornsson G., Andersen M.I. et al. 1999, Science 283,
       2073
\reference di Serego Alighieri, S. 1997, in Instrumentation for Large
        Telescopes, ed. J. M. Rodriguez Espinosa, A. Herrero, \& F. Sanchez
(Cambridge
        University Press), 287
% \reference Ghisellini G. \& Celotti A. 1999, ApJ 511, L93
\reference Ghisellini G. \& Lazzati D. 1999, MNRAS 309, L17
\reference Granot J., Panaitescu A., Kumar P., Woosley S.E. 2002, ApJ 570, L61
\reference Gruzinov A. 1999, ApJ 525, L29
\reference Gruzinov A. \& Waxman E. 1999, ApJ 511, 852
\reference Klose S., Stecklum B., Fischer O. 2001, Gamma--Ray Bursts in the
         Afterglow Era, eds. E. Costa, F. Frontera, J. Hjorth (Springer), 188
\reference Loeb A. \& Perna R. 1998, ApJ 495, 597
\reference Masetti N., Palazzi E., Pian E. et al. 2003, this proceedings
% \reference Meszaros P. \& Rees M. 1997, ApJ 476, 232
\reference Medvedev M.V. \& Loeb A. 1999, ApJ 526, 697
\reference Rol E., Wijers R.A.M.J., Vreeswijk P.M. et al. 2000, ApJ 544, 707
\reference Rol E.,  Castro Cer\'on J.M., Gorosabel J. et al. 2002, GCN\,1596
\reference Rol E., Gorosabel J., Palazzi E. et al. 2003, this proceeding
\reference Rossi E.M., Lazzati D. \& Rees M.J., 2002, MNRAS, 332, 945
\reference Rossi E.M., Lazzati D., Salmonson J.D. \& Ghisellini G. 2002,
    Beaming and Jets in Gamma Ray Bursts, Copenhagen (astro-ph/0211020)
\reference Sari R. 1999, ApJ 524, L43
\reference Wang L., Baade D., H\"oflich P. \& Wheeler J.C. 2003, A\&AL,
           submitted (astro-ph/0301266)
\reference Wijers R.A.M.J., Vreeswijk P. M., Galama T.J. et al. 1999, ApJ 523,
           L33
\reference Zhang B. \& M\'esz\'aros P., 2002, ApJ, 571, 876
\end{references}
\end{document}